\documentclass[aps,prc,preprint,amsmath,showpacs,superscriptaddress,nofootinbib]{revtex4-1}
\usepackage{graphicx,epstopdf,epsfig,psfrag}
\newcommand{\beqy}{\begin{eqnarray}}
\newcommand{\eeqy}{\end{eqnarray}}
\newcommand{\bmlet}{\begin{subequations}}
\newcommand{\emlet}{\end{subequations}}

\begin{document}

\title{Analytical determination of the structure of the outer crust of a cold nonaccreted neutron star: extension to strongly quantizing magnetic fields}

\author{N. Chamel}
\affiliation{Institut d'Astronomie et d'Astrophysique, CP-226, Universit\'e Libre de Bruxelles, 
1050 Brussels, Belgium}
\author{Zh. K. Stoyanov}
\affiliation{Institute for Nuclear Research and Nuclear Energy, Bulgarian Academy of Sciences, 72 Tsarigradsko Chaussee, 1784 Sofia, Bulgaria}

\begin{abstract}
The iterative method recently proposed for determining the internal constitution of the outer crust of a nonaccreted neutron star is extended to magnetars by taking into account the Landau-Rabi quantization of electron motion induced by the presence of a very high magnetic field. 
It is shown that in the strongly quantizing regime, the method can be efficiently implemented using new analytical solutions for the transitions between adjacent crustal layers. Detailed numerical computations are performed to assess the performance and precision of the method. 
\end{abstract}


\maketitle

\section{Introduction}

Although most neutron stars are endowed with typical magnetic fields of order $10^{12}$~G, the subclass of magnetars~\cite{thompson1992} - including anomalous X-ray pulsars and soft gamma-ray repeaters (SGR) - exhibit much higher fields, up to a few times $10^{15}$~G at their surface~\cite{turolla2015,kaspi2017}.  Potentially even more extreme magnetic fields could be sustained in their interior, as shown by numerical simulations~\cite{uryu2019}. 
Giant flares, such as those observed in SGR 1806$-$20, are among the most spectacular astrophysical manifestations of the magnetic activity, whereby sudden changes in the magnetic field configuration are accompanied by starquakes, as suggested by the detection of quasiperiodic oscillations (see, e.g. Ref.~\cite{glampedakis2018} for a recent review). The frequencies of the various modes depend on the internal constitution of these stars. However, the identification of these modes remains challenging due to uncertainties on the stellar structure, in particular on the properties of the crust~\cite{nandi2016,tews2017,sotani2018}. Some parts of the crust may actually be ejected during such events~\cite{gelfand2007}. The subsequent decompression of this neutron-rich material provides suitable conditions for the rapid neutron capture process so called $r$-process at the origin of stable and some long-lived radioactive neutron-rich nuclides heavier than iron~\cite{arnould2007}. The final nuclear abundances of the processed stellar material depend on the initial composition of magnetar crusts. The crustal properties are also important for the long-term evolution of the magnetic field and the cooling of the star~\cite{mereghetti2015,potekhin2015,pons2019}. 

The internal structure of a neutron star can be significantly altered by the presence of a high 
magnetic field, especially in the crust region (see, e.g., Ref.~\cite{blaschke2018} for a recent review). The composition of the outer crust of a magnetar has been traditionally determined following the study of Ref.~\cite{lai91b} by minimizing the Gibbs free energy per nucleon $g$ at zero temperature and for a finite set of pressure values (see, e.g. Refs.~\cite{nandi2011,chapav2012,basilico2015,mutafchieva2019}). The pressure step must be small enough to find the complete stratification, especially in the deepest region of the outer crust where even a thin layer can contain the most abundant nuclear species. Systematic calculations over a wide range of magnetic-field strengths, as required for the modelling of magnetars, can thus become computationally very expensive. 

In this paper, the computationally very fast approach recently proposed to calculate 
the structure of the outer crust of unmagnetized neutron stars~\cite{chamel2020,chamel2020b} is extended to take into account the presence of a strongly quantizing magnetic field. New analytical solutions for the transition pressure between adjacent crustal layers are presented in Sect.~\ref{sec:transition}. The analytical approximations for the nuclear abundances and the depth of the different layers that were previously discussed in Ref.~\cite{chamel2020} are suitably generalized to magnetars in Sect.~\ref{sec:structure}. The numerical implementation of all these formulas is discussed in Sect.~\ref{sec:implementation}, where numerical tests of their precision are also presented. 

\section{Transition between adjacent crustal layers}
\label{sec:transition}

In the following, we shall consider the crustal region at densities $\rho$ above the ionization threshold and below the neutron-drip point. As in Ref.~\cite{chamel2020}, we assume that each crustal layer is made of a single nuclear species ($A$, $Z$) with mass number $A$ and atomic number $Z$ in thermodynamic equilibrium at temperature $T$ below the crystallization temperature $T_m$ (for all practical purposes, we shall set $T=0$~K). 

The equilibrium composition at pressure $P$ is determined by the minimization of the Gibbs free energy per nucleon given by (see, e.g. Ref.~\cite{chapav2012})
\begin{equation}
\label{eq:gibbs}
g(A,Z,n_e) = \frac{M^\prime(A,Z)c^2}{A} + \frac{Z}{A}\bigg[\mu_e-m_e c^2+\frac{4}{3}C \alpha \hbar c  n_e^{1/3} Z^{2/3} \biggr]\, ,
\end{equation}
where $M^\prime(A,Z)$ is the mass of the nucleus ($A$,$Z$) (including the rest mass of $Z$ electrons), $m_e$ is the electron mass, $\mu_e$ is the Fermi energy of a relativistic electron gas with number density $n_e$, $C$ is the crystal lattice structure constant, and $\alpha=e^2/(\hbar c)$ is the fine structure constant ($e$ being the elementary electric charge, $\hbar$ the Planck-Dirac constant and $c$ the speed of light). The pressure is expressible in terms of the electron number density $n_e$ by the relation 
\begin{equation}
\label{eq:pressure}
P=P_e + C\, \alpha \hbar c Z^{2/3} n_e^{4/3}\, , 
\end{equation}
where $P_e$ denotes the pressure of an ideal electron Fermi gas (see, e.g. Ref.~\cite{haensel2007} for general expressions). To first order in $\alpha$, the transition from a crustal layer made of nuclei ($A_1$, $Z_1$) to 
a denser layer made of nuclei ($A_2$, $Z_2$) is formally determined by the same condition as in the absence of magnetic fields, and is approximately given by~\cite{chamel2017a}: 
\begin{equation}\label{eq:threshold-condition}
\mu_e + C\, \alpha \hbar c n_e^{1/3}F(Z_1,A_1 ; Z_2, A_2) = \mu_e^{1\rightarrow 2}\, ,
\end{equation}
\begin{equation}\label{eq:def-F}
F(Z_1, A_1 ; Z_2, A_2)\equiv \left(\frac{4}{3}\frac{Z_1^{5/3}}{A_1} - \frac{1}{3}\frac{Z_1^{2/3} Z_2}{A_2} -\frac{Z_2^{5/3}}{A_2}\right)
\left(\frac{Z_1}{A_1}-\frac{Z_2}{A_2}\right)^{-1} \, ,
\end{equation}
\begin{equation}\label{eq:muethres}
\mu_e^{1\rightarrow 2}\equiv \biggl[\frac{M^\prime(A_2,Z_2) c^2}{A_2}-\frac{M^\prime(A_1,Z_1)c^2}{A_1}\biggr]\left(\frac{Z_1}{A_1}-\frac{Z_2}{A_2}\right)^{-1} +  m_e c^2\, .
\end{equation}
The singular case $Z_1/A_1=Z_2/A_2$ needs not be considered as it leads to much higher densities than any other transition (see, e.g., the discussion in Appendix A of Ref.~\cite{chamel2016}). 
The baryon chemical potential $\mu_{1\rightarrow2}$ and the pressure $P_{1\rightarrow2}$ at the interface between the two layers both vary continuously and can thus  be calculated from Eqs.~(\ref{eq:gibbs}) and (\ref{eq:pressure})  respectively, with $Z=Z_1$, $A=A_1$. 
The transition is
generally accompanied by a discontinuous change of the mean nucleon number density: 
\begin{equation}
\bar n_1^{\rm max} = \frac{A_1}{Z_1} n_e\, , 
\end{equation}
\begin{equation}
\bar n_2^{\rm min} = \frac{A_2}{Z_2} n_e \Biggl[ 1+\frac{1}{3}C \alpha \hbar c n_e^{1/3} (Z_1^{2/3}-Z_2^{2/3})\left(\frac{dP_e}{dn_e}\right)^{-1} \Biggr]\, .
\end{equation}
The bottom of the outer crust is marked by the onset of neutron emission by nuclei. This process is determined by the following equations~\cite{chamel2015b}
\begin{equation}\label{eq:n-drip-mue}
\mu_e + \frac{4}{3}C \alpha \hbar c n_e^{1/3} Z^{2/3} =  \mu_e^{\rm drip}\, , 
\end{equation}
\begin{equation}\label{eq:muedrip}
\mu_e^{\rm drip}\equiv \frac{-M^\prime(A,Z)c^2+A m_n c^2}{Z} +m_e c^2 \, ,
\end{equation}
where $m_n$ is the neutron mass.

In the presence of a magnetic field, the electron motion perpendicular to the field is quantized, as first shown by Rabi~\cite{rabi1928}. The magnetic field is strongly quantizing if $B_\star\equiv B/B_{\rm rel}\gg 1$ with 
\begin{equation}
\label{eq:Bcrit}
B_\textrm{rel}=\left(\frac{m_e c^2}{\alpha \lambda_e^3}\right)^{1/2}\approx 4.4\times 10^{13}\, \rm G\, , 
\end{equation}
where $\lambda_e=\hbar/(m_e c)$ is the electron Compton wavelength. 
For a given magnetic field strength $B_\star$, the number of occupied levels is determined by the condition
\begin{equation}
n_e =\frac{2 B_\star}{(2 \pi)^2 \lambda_e^3} \sum_{\nu=0}^{\nu_{\rm  max}} g_\nu x_e(\nu)\, ,
\end{equation}
\begin{equation}
x_e(\nu) =\sqrt{\gamma_e^2 -1-2 \nu B_\star}\, ,
\end{equation}
where $\gamma_e=\mu_e/(m_e c^2)$, $g_\nu=1$ for $\nu=0$ and $g_\nu=2$ for $\nu\geq 1$. For a given value of the Fermi energy $\mu_e$, the electron number density $n_e$ exhibits typical quantum oscillations as a function of $B_\star$.

For $B_\star\geq (\gamma_e^2-1)/2$, electrons are confined to the lowest Rabi level. The equilibrium condition~(\ref{eq:threshold-condition}) is amenable to analytical solutions if the electron density in the second term of the left hand side is expressed in terms of the electron Fermi energy using the ultrarelativistic approximation 
\begin{equation}\label{eq:mue-strongB}
n_e  \approx \frac{B_\star} {2 \pi^2 \lambda_e^3 }\gamma_e\, .
\end{equation}
Introducing 
\begin{equation}
\bar F(Z_1,A_1;Z_2,A_2; B_\star)\equiv \frac{1}{3} C\alpha F(Z_1,A_1;Z_2,A_2)\left(\frac{B_\star}{2\pi^2}\right)^{1/3}\, ,
\end{equation}
Eq.~(\ref{eq:threshold-condition}) thus reduces to 
\begin{equation}\label{eq:threshold-condition-strongB}
\gamma_e + 3\bar F(Z_1,A_1;Z_2,A_2; B_\star) \gamma_e^{1/3} = \gamma_e^{1\rightarrow 2}\, ,
\end{equation}
which can be expressed as a cubic polynomial equation. Introducing the dimensionless parameter 
\begin{equation}
\upsilon\equiv \frac{\gamma_e^{1\rightarrow 2}}{2 |\bar F(Z_1,A_1;Z_2,A_2; B_\star)|^{3/2}}\, ,
\end{equation}
and using the known analytical expressions for the real roots of cubic equations  (see, e.g., Ref.~\cite{birkhoff2010}), the solutions of Eq.~(\ref{eq:threshold-condition-strongB}) for $\gamma_e$ are given by the following formulas: 
\begin{itemize}
	\item $\bar F(Z_1,A_1;Z_2,A_2; B_\star)>0$
	\begin{equation}
	\gamma_e=8\bar F(Z_1,A_1;Z_2,A_2; B_\star)^{3/2}\, {\rm sinh}^3 \left(\frac{1}{3}{\rm arcsinh~} \upsilon\right)\, ,
	\end{equation}
	\item $\bar F(Z_1,A_1;Z_2,A_2; B_\star)<0$
	\begin{equation}
	\gamma_e=\begin{cases}
	8|\bar F(Z_1,A_1;Z_2,A_2; B_\star)|^{3/2}\, {\rm cosh}^3\left(\frac{1}{3}{\rm arccosh~} \upsilon\right) & \text{if} \ \upsilon\geq 1\, ,\\
	8|\bar F(Z_1,A_1;Z_2,A_2; B_\star)|^{3/2}\, \cos^3\theta_k & \text{if} \ 0\leq \upsilon< 1\, . 
	\end{cases}
	\end{equation}
	with
	\begin{equation}
	\theta_k\equiv \frac{1}{3}\arccos \upsilon + \frac{2\pi k}{3}\, ,
	\end{equation}
	and $k=0,1,2$. 
\end{itemize}
The mathematical solutions $k=1$ and $k=2$ yield $\gamma_e \leq 0$, and they must therefore be discarded. 
The transition pressure and the densities of the layers are given by 
\begin{equation}\label{eq:P-strongB}
P_{1\rightarrow 2} = \frac{B_\star m_e c^2 }{4 \pi^2 \lambda_e^3 }\biggl[x_e+
\sqrt{1+x_e^2}-\log\left(x_e+\sqrt{1+x_e^2}\right)
+\left(\frac{4 B_\star Z_1^2 x_e^4}{\pi^2}\right)^{1/3}\frac{C \alpha }{3}\biggr] \, ,
\end{equation} 
\begin{equation}\label{eq:n1max-strongB}
\bar n_1^{\rm max} = \frac{B_\star}{2\pi^2 \lambda_e^3} \frac{A_1}{Z_1} x_e
\end{equation}
\begin{equation}\label{eq:n2min-strongB}
\bar n_2^{\rm min} = \frac{B_\star}{2\pi^2 \lambda_e^3} \frac{A_2}{Z_2} x_e\biggl[1+ \frac{1}{3}C \alpha (Z_1^{2/3}-Z_2^{2/3})  \left(\frac{B_\star}{2\pi^2}\right)^{1/3}\frac{\sqrt{1+x_e^2}}{x_e^{5/3}}\biggr]\, ,
\end{equation}
where $x_e=\sqrt{\gamma_e^2-1}$. 

For high enough magnetic fields, the second term in the left-hand side of Eq.~(\ref{eq:threshold-condition-strongB}) can be larger than $\gamma_e$ so that the equilibrium composition corresponds to $\gamma_e^{1\rightarrow 2}<0$. Although the expansion of the Gibbs free energy is not expected to be accurate in this case, analytical solutions can still be of interest as a first initial guess in the search for the numerical value of $\gamma_e$. Real solutions only exist for $\bar F(Z_1,A_1;Z_2,A_2; B_\star)<0$ and $-1<\upsilon\leq 0$:   
\begin{equation}
\gamma_e=8|\bar F(Z_1,A_1;Z_2,A_2; B_\star)|^{3/2}\, \cos^3\theta_k \, ,
\end{equation}
and $k=0,1,2$. 
As in the case of ``low'' magnetic fields (but still strongly quantizing), the solution $k=1$ must be ignored since $\gamma_e<0$. However, both $k=0$ and $k=2$ now leads to $\gamma_e\geq 0$. The physically admissible solution is determined by selecting the expression yielding the lowest transition pressure satisfying the conditions $\gamma_e\geq 1$ and $\bar n_2^{\rm min}\geq \bar n_1^{\rm max}$, as required by mechanical stability. 
Solutions for the neutron-drip transition can be found using the above formulas after substituting $F(Z_1,A_1;Z_2,A_2)$ by $(4/3) Z^{2/3}$ and $\gamma_e^{1\rightarrow 2}$ by $\gamma_e^{\rm drip}$. 

\section{Global structure and nuclear abundances}
\label{sec:structure}

In principle, the global structure of a highly-magnetized neutron star should be calculated solving simultaneously Einstein's and Maxwell's equations. However, 
the influence of the magnetic field on the crust size was shown to lie below about $1-2\%$ for $B_\star \lesssim 10^4$~\cite{franzon2017,gomes2019, chatterjee2019,chatterjee2019b}. We shall thus employ the same analytical formulas as those derived for unmagnetized neutron stars in Ref.~\cite{chamel2020}. 
The relative nuclear abundance of a crustal layer  is thus approximately given by 
\begin{equation}\label{eq:xi}
\xi= \frac{\delta P}{P_{\rm drip}}\, ,
\end{equation}
where $\delta P$ the range of pressures of the layer under consideration and $P_{\rm drip}$ 
is the neutron-drip pressure. The associated baryonic mass for a star with a gravitational 
mass $\mathcal{M}$ and a circumferential radius $R$ can be obtained as follows: 
\begin{equation}\label{eq:baryonic-mass}
\delta\,M_B \approx \xi \frac{8\pi R^4 P_{\rm drip}}{r_g c^2} \left(1-\frac{r_g}{R}\right)^{3/2}  \, ,
\end{equation}
where $r_g=2 G\mathcal{M}/c^2$ is the Schwarzschild radius. 
The proper depth $z$ below the surface at the transition between two adjacent crustal layers with baryon chemical potential $\mu_{1\rightarrow 2}$ is 
approximately given by 
\begin{equation}\label{eq:z}
z \approx z_{\rm drip} \frac{(\mu_{1\rightarrow 2}/\mu_s)^2-1}{(m_n c^2/\mu_s)^2-1} \, ,
\end{equation}
where $\mu_s$ is the baryon chemical potential at the stellar surface (where $P=0$), and the depth at the neutron-drip transition is given by 
\begin{equation}\label{eq:zdrip}
z_{\rm drip}\approx \frac{R^2}{r_g}\biggl[\left(\frac{m_n c^2}{\mu_s}\right)^2-1\biggr]\sqrt{1-\frac{r_g}{R}}\, .
\end{equation}
Contrary to the case of unmagnetized neutron stars, $\mu_s/c^2$ is not simply given by the mass  $m_0$ per nucleon of $^{56}$Fe because the density at the surface is finite and is approximately given by~\cite{lai91b,chapav2012} 
\begin{equation}
\label{eq:ns}
n_s \approx \frac{A_s}{\lambda_e^3}\biggl[\frac{|C|  \alpha B_\star^2 }{4\pi^4 Z_s}\biggr]^{3/5} \, ,
\end{equation}
with $Z_s=26$ and $A_s=56$ the corresponding atomic and mass numbers of $^{56}$Fe. The corresponding value of $\mu_s$ can be calculated from Eq.~(\ref{eq:gibbs}) with $n_e=(Z_s/A_s)n_s$.

\section{Stratification of the outer crust}
\label{sec:implementation}

The stratification of the outer crust is determined as in the case of unmagnetized neutron stars~\cite{chamel2020}. Given a crustal layer made of nuclide ($A_1$, $Z_1$), the composition of the layer beneath can be found by merely determining the nuclide ($A_2$, $Z_2$) leading to the lowest transition pressure $P_{1\rightarrow 2}$. Starting with $^{56}$Fe at the stellar surface, the sequence of equilibrium nuclides can thus be determined iteratively. The iteration is stopped when the baryon chemical potential exceeds the neutron mass energy. Once the composition has been found, the detailed structure of the crust and the nuclear abundances can be readily calculated using the analytical formulas for the pressure and baryon chemical potential at the interface between adjacent layers. 

To assess the efficiency of 
the method in the strongly quantizing regime, we have calculated the internal constitution of the outer crust of a nonaccreted magnetar with $B_\star=2000$ using experimental data from the 2016 Atomic Mass Evaluation~\cite{ame2016} supplemented with the nuclear mass table HFB-27 from the {\footnotesize BRUSLIB} database~\cite{bruslib} and based on the Hartree-Fock-Bogoliubov method~\cite{hfb27}. We have also made use of the recent measurements of copper isotopes~\cite{welker2017}. Nuclear masses were estimated from tabulated \emph{atomic} masses after subtracting out the electron binding energy using Eq.~(A4) of Ref.~\cite{lunney2003}. 
In each layer, nuclei are arranged in a body-centered cubic lattice independently of the magnetic field strength~\cite{kozhberov2016}. The structure constant is taken from Ref.~\cite{baiko2001}. 
 Results are summarized in Table~\ref{tab1}. The computations took about 0.07 seconds using an Intel Core i7-975 processor. In contrast, the standard approach using about 19000 different pressure values between $P=9\times 10^{-12}$ MeV~fm$^{-3}$ and $P=P_{\rm drip}$ with a pressure step $\delta P=10^{-3}P$ took about 24 minutes, i.e. $\approx 2\times 10^4$ times longer. Comparing with the results obtained in Ref.~\cite{chamel2020}, the magnetic field changes the composition of the crust: the layers made of $^{64}$Ni, $^{66}$Ni, and $^{78}$Ni have disappeared, while new layers made of nuclei $^{88}$Sr, $^{132}$Sn, $^{128}$Pd are now present. Due to the increase of the matter density induced by the magnetic field, matter is more uniformly distributed: the baryonic content of the shallow layers is now comparable to that of the deeper layers. In these calculations, the same nuclear masses as in the absence of magnetic fields were employed. However, high enough magnetic fields can also influence the structure of nuclei~\cite{arteaga2011,stein2016}, inducing additional changes in the crustal composition~\cite{basilico2015}. 
 
 We have determined the precision of the method by solving
 numerically the exact equilibrium conditions: 
\begin{equation}
\label{eq:equilibrium1}
g(A_1,Z_1,n_e^1)=g(A_2,Z_2,n_e^2)\equiv \mu_{1\rightarrow2}\, ,
\end{equation}
\begin{equation}
\label{eq:equilibrium2}
P(n_e^1,Z_1)=P(n_e^2,Z_2)\equiv P_{1\rightarrow2}\, .
\end{equation}
The relative deviations between these 
results and the analytical formulas are indicated in Table~\ref{tab2}. In most cases, the errors on the pressures and densities do not exceed 0.24\%. The errors of a few \% found for the transition from $^{56}$Fe to $^{62}$Ni in the shallow region of the crust where electrons are only moderately relativistic (as indicated by the rather low value of the parameter $x_e$) can be traced back to the approximation~(\ref{eq:mue-strongB}). The transition from $^{132}$Sn to $^{80}$Zn also exhibits comparatively large deviations, but their origin is different. As indicated in Table~\ref{tab2}, the threshold electron chemical potential $\mu_e^{1\rightarrow 2}$ associated with this transition is negative so that the second term in Eq.~(\ref{eq:threshold-condition}) must be large and negative. However, the expansion of the Gibbs free energy per nucleon to first order in $\alpha$ requires this term to be small. 
Except for the two peculiar cases discussed above, the depths are determined with an error of 0.14~\% at most. As expected, the relative abundances being obtained from pressure differences exhibit larger deviations, especially in the vicinity of the transitions from $^{56}$Fe to $^{62}$Ni  and from $^{132}$Sn to $^{80}$Zn. On the other hand, the analytical formulas for the baryon chemical potentials remain very accurate in all cases, with deviations below $8\times 10^{-3}$\%, thus ensuring that the sequence of equilibrium nuclides is correctly reproduced. Having found the composition, the crustal properties could thus be refined in a second stage by solving numerically Eqs.~(\ref{eq:equilibrium1}) and (\ref{eq:equilibrium2}). The overall procedure will still remain much faster than the full minimization.

\section{Conclusions}

We have extended the iterative method proposed in Ref.~\cite{chamel2020} for determining the structure and the composition of the outer crust of a cold nonaccreted neutron star to allow for the Landau-Rabi quantization of the electron motion induced by the presence of a magnetic field. We have shown that this method can be very efficiently implemented  in the limit of a strongly quantizing magnetic field by making use of new analytical solutions for the transitions between adjacent crustal layers. Computations are found to be as fast as for unmagnetized neutron stars. Computer codes have been made publicly available for both unmagnetized~\cite{chamel2020b} and strongly magnetized neutron stars~\cite{chamel2020c}. The general scheme proposed in Ref.~\cite{chamel2020} is therefore particularly well-suited for systematic calculations of the equation of state of dense magnetized matter for a large number of different magnetic-field strengths, as required for the modelling of magnetars.

\begin{table}
	\centering
	\caption{Stratification of the outer crust of a magnetar with $B_\star=2000$, as obtained using recent experimental data supplemented with the nuclear mass model HFB-27~\cite{hfb27}. In the table are listed: the atomic numbers $Z_1$ and $Z_2$ of adjacent layers, the corresponding mass numbers $A_1$ and $A_2$, the dimensionless parameter $x_e$, the maximum and minimum mean nucleon number densities $\bar{n}_1^{\rm max}$ and $\bar{n}_2^{\rm min}$ at which the nuclides are  present, the transition pressure 
		$P_{1\rightarrow 2}$, the electron and baryon threshold chemical potentials $\mu_e^{1\rightarrow 2}$ and $\mu_{1\rightarrow 2}$, the relative abundance $\xi_1$ of nuclide ($A_1$, $Z_1$) and its relative depth $z_1/z_{\rm drip}$. Units are megaelectronvolts for energy and femtometers for length. See text for details.}
	\label{tab1}
	\vspace{.5cm}
	\begin{tabular}{|cccccccccccc|}
		\hline
		$Z_1$ & $A_1$ & $Z_2$ & $A_2$   &   $x_e$      &      $\bar n_1^{\rm max}$ &    $\bar n_2^{\rm min}$   &    $P_{1\rightarrow 2}$  &   $\mu_e^{1\rightarrow 2}$   &     $\mu_{1\rightarrow 2}$   &      $\xi_1$           &     $z_1/z_{\rm drip}$ \\
		\hline
26 & 56 & 28 &62   &   1.31   &     4.98$\times 10^{-6}$ & 5.15$\times 10^{-6}$ &   3.00$\times 10^{-7}$ &   0.966   & 930.4  &    2.62$\times 10^{-4}$ &   8.67$\times 10^{-3}$  \\
28 & 62 & 38 &88   &   5.68   &     2.21$\times 10^{-5}$ & 2.34$\times 10^{-5}$ &   1.23$\times 10^{-5}$ &   4.44    & 931.3  &    1.04$\times 10^{-2}$ &   0.101             \\
38 & 88 & 36 &86   &   8.26   &     3.37$\times 10^{-5}$ & 3.47$\times 10^{-5}$ &   2.69$\times 10^{-5}$ &   2.84    & 931.8  &    1.28$\times 10^{-2}$ &   0.156             \\
36 & 86 & 34 &84   &   13.1  &     5.49$\times 10^{-5}$ &  5.67$\times 10^{-5}$ &   7.06$\times 10^{-5}$ &   5.13     & 932.8  &   3.82$\times 10^{-2}$ &   0.261             \\
34 & 84 & 32 &82   &   18.6  &     8.08$\times 10^{-5}$ &  8.37$\times 10^{-5}$ &   1.46$\times 10^{-4}$ &   7.83     & 933.9  &   6.60$\times 10^{-2}$ &   0.380             \\
32 & 82 & 50 &132  &   23.9  &     1.08$\times 10^{-4}$ &  1.12$\times 10^{-4}$ &   2.44$\times 10^{-4}$ &   19.6     & 934.9  &   8.55$\times 10^{-2}$ &   0.491             \\
50 & 132& 30 &80  &   25.1  &     1.17$\times 10^{-4}$ &   1.17$\times 10^{-4}$ &   2.67$\times 10^{-4}$ &   -17.0     & 935.1  &  1.98$\times 10^{-2}$ &   0.513             \\
30 & 80 & 46 &128  &  28.2  &     1.32$\times 10^{-4}$ &  1.39$\times 10^{-4}$  &   3.44$\times 10^{-4}$ &   19.0     & 935.7  &   6.72$\times 10^{-2}$ &   0.579            \\
46 & 128& 44 &126 &   34.5  &     1.69$\times 10^{-4}$ &   1.74$\times 10^{-4}$ &   5.11$\times 10^{-4}$ &   15.2      & 936.8  &  0.146                &   0.697             \\
44 & 126& 42 &124 &   37.8  &     1.91$\times 10^{-4}$ &   1.96$\times 10^{-4}$ &   6.18$\times 10^{-4}$ &   16.9      & 937.4  &  9.39$\times 10^{-2}$ &   0.761             \\
42 & 124& 40 &122 &   42.8  &     2.22$\times 10^{-4}$ &   2.29$\times 10^{-4}$ &   7.95$\times 10^{-4}$ &   19.4      & 938.2  &  0.154                &   0.853             \\
40 & 122& 38 &120 &   45.2  &     2.43$\times 10^{-4}$ &   2.51$\times 10^{-4}$ &   8.90$\times 10^{-4}$ &   20.7      & 938.6  &  8.36$\times 10^{-2}$ &   0.897             \\
38 & 120& 38 &122 &   50.0  &     2.78$\times 10^{-4}$ &   2.82$\times 10^{-4}$ &   1.09$\times 10^{-3}$ &   24.2      & 939.4  &  0.175                &   0.980             \\
38 & 122& 38 &124 &   51.0  &     2.88$\times 10^{-4}$ &   2.93$\times 10^{-4}$ &   1.14$\times 10^{-3}$ &   24.7      & 939.5  &  4.11$\times 10^{-2}$ &   0.998             \\
38 & 124& $-$  &$-$&  51.2  &     2.94$\times 10^{-4}$ &                  $-$   &   1.14$\times 10^{-3}$ &   24.8      & 939.6  &  5.47$\times 10^{-3}$ &   1.00       \\
\hline
	\end{tabular}
\end{table}

\begin{table}
	\centering
	\caption{Precision of the calculated properties of the outer crust of a magnetar, as listed in Table~\ref{tab1}. The relative deviation $\delta q$ (in \%) of a quantity $q$ is calculated 
		as $\delta q=100(q-q_\textrm{exact})/q_\textrm{exact}$, where $q_\textrm{exact}$ is the exact value while $q$ denotes the value calculated using the analytical formulas. See text for details. }
	\label{tab2}
	\vspace{.5cm}   
	\begin{tabular}{|ccccccccccc|}
		\hline
		$Z_1$ & $A_1$ & $Z_2$ & $A_2$   &   $x_e$      &      $\bar n_1^{\rm max}$ &    $\bar n_2^{\rm min}$   &    $P_{1\rightarrow 2}$  &    $\mu_{1\rightarrow 2}$   &      $\xi_1$           &     $z_1/z_{\rm drip}$ \\
		\hline
		26 & 56 & 28 & 62   & -1.4 & -1.4 &  -1.6 &  -4.6  & -3.1$\times 10^{-4}$ &  -4.6 &  -3.5  \\
		28 & 62 & 38 & 88   & -1.1$\times 10^{-1}$& -1.1$\times 10^{-1}$ & -1.9$\times 10^{-1}$ &  -2.4$\times 10^{-1}$ & -1.4$\times 10^{-4}$ &  -1.2$\times 10^{-1}$ &  -1.4$\times 10^{-1}$  \\
		38 & 88 & 36 & 86   & 9.7$\times 10^{-2}$ & 9.7$\times 10^{-2}$ & 1.1$\times 10^{-1}$ & 2.1$\times 10^{-1}$  & 1.8$\times 10^{-4}$ & 5.8$\times 10^{-1}$  & 1.1$\times 10^{-1}$ \\
		36 & 66 & 34 & 84   & 2.7$\times 10^{-2}$ & 2.7$\times 10^{-2}$ & 3.3$\times 10^{-2}$ & 5.7$\times 10^{-2}$  & 7.8$\times 10^{-5}$ & -3.6$\times 10^{-2}$ & 3.0$\times 10^{-2}$  \\
		34 & 84 & 32 & 82   & 1.0$\times 10^{-2}$ & 1.0$\times 10^{-2}$ & 1.4$\times 10^{-2}$ & 2.1$\times 10^{-2}$  & 4.1$\times 10^{-5}$ & -1.2$\times 10^{-2}$ & 1.1$\times 10^{-2}$ \\
		32 & 82 & 50 & 132  & 5.3$\times 10^{-2}$ & 5.3$\times 10^{-2}$ & 3.0$\times 10^{-2}$ & 1.1$\times 10^{-1}$  & 2.6$\times 10^{-4}$ & 2.4$\times 10^{-1}$  & 5.4$\times 10^{-2}$  \\
		50 & 132 & 30 & 80  & 1.6                 & 1.6                 & 1.7                 & 3.3                  & 8.0$\times 10^{-3}$   & 5.9$\times 10^1$     & 1.6   \\
		30 & 80 & 46 & 128  & 2.7$\times 10^{-2}$ & 2.7$\times 10^{-2}$ & 1.2$\times 10^{-2}$ & 5.6$\times 10^{-2}$  & 1.6$\times 10^{-4}$ & -9.9                 & 2.7$\times 10^{-2}$  \\
		46 & 128 & 44 & 126 & 2.6$\times 10^{-3}$ & 2.6$\times 10^{-3}$ & 4.2$\times 10^{-3}$ & 5.4$\times 10^{-3}$  & 1.8$\times 10^{-5}$ & -1.0$\times 10^{-1}$ &  2.6$\times 10^{-3}$ \\
		44 & 126&  42&  124 & 2.1$\times 10^{-3}$ & 2.1$\times 10^{-3}$ & 3.4$\times 10^{-3}$ & 4.2$\times 10^{-3}$  & 1.5$\times 10^{-5}$ & -2.5$\times 10^{-3}$ & 1.9$\times 10^{-3}$  \\
		42 & 124&  40&  122 & 1.5$\times 10^{-3}$ & 1.5$\times 10^{-3}$ & 2.6$\times 10^{-3}$ & 3.0$\times 10^{-3}$  & 1.1$\times 10^{-5}$ & -1.8$\times 10^{-3}$ & 1.4$\times 10^{-3}$ \\
		40 & 122&  38&  120 & 1.3$\times 10^{-3}$ & 1.3$\times 10^{-3}$ & 2.3$\times 10^{-3}$ & 2.5$\times 10^{-3}$  & 9.9$\times 10^{-6}$ & -2.0$\times 10^{-3}$ & 1.1$\times 10^{-3}$ \\
		38 & 120&  38&  122 & 3.7$\times 10^{-4}$& 3.7$\times 10^{-4}$& 3.7$\times 10^{-4}$& 7.5$\times 10^{-4}$ & 3.1$\times 10^{-6}$     & -7.9$\times 10^{-3}$ & 3.3$\times 10^{-4}$ \\
		38 & 122&  38&  124 & 3.5$\times 10^{-4}$& 3.5$\times 10^{-4}$& 3.5$\times 10^{-4}$& 7.1$\times 10^{-4}$ & 3.0$\times 10^{-6}$     & -9.4$\times 10^{-4}$ & 3.1$\times 10^{-4}$ \\
		38 & 124&$-$ &  $-$ & 3.5$\times 10^{-4}$ & 3.5$\times 10^{-4}$ &    $-$              & 7.0$\times 10^{-4}$  & $-$                 & -9.4$\times 10^{-4}$ &  $-$  \\ 
		\hline
	\end{tabular}
\end{table}

\section*{Acknowledgments}
This work of N.C. was financially supported by Fonds de la Recherche Scientifique (Belgium) under grant no. IISN 4.4502.19. The work of Z. S. was financially supported by the National programme ``Young scientists'' funded by the Bulgarian Ministry of Education and Science and a Short Term Scientific Mission grant from the European Cooperation in Science and Technology (COST) Action CA16117. This work was also partially supported by the COST Action CA16214.

\end{document}